\documentclass[12pt]{iopart}
\usepackage{graphicx}

\begin{document}

\title[]{Stability, Adiabaticity and Transfer efficiency
in a nonlinear $\Lambda$-system\\ }

\author{Ning Jia, Jing Qian$^{\dagger}$, Guangjiong Dong and Weiping Zhang}

\address{Department of Physics, Quantum Institute for Light and Atoms, East China Normal University, Shanghai 200062, People's Republic of China}
\ead{jqian1982@gmail.com}
\begin{abstract}
We investigate the relationship between stability, adiabaticity and transfer
efficiency in a $\Lambda $-type atom-molecule coupling system via a nonlinear
stimulated Raman adiabatic passage. We find that only when the pump and control lasers overlap in time domain, the coherent population trapping (CPT) state could become unstable. If the overlapping time of the two lasers is short so that unstable growth of the deviation from the CPT state is negligible, then good adiabaticity of the CPT state could be maintained even in the unstable region. In this case, a high atom-molecule transfer efficiency could be obtained  by chirping applied laser pulses to elegantly compensate the frequency shift induced by intra-atomic collision. Our results could be useful for efficiently photoassociating ground-state molecules from a cold atomic gas with strong atom-atom collisional interaction.



\end{abstract}

\submitto{\JPB}
\maketitle

\section{Introduction}

Motivated by the success of cold atomic physics, recently, there has been a growing interest in producing cold and ultracold molecules in molecular physics \cite{Doyle04,Carr09}. Cold molecules could be produced by sympathetic cooling \cite{Lara06} or deceleration of supersonic beams \cite{Hendrick02}. These methods are general; however, it is still hard to attain ultracold  molecules($<$1mK). Despite the difficulty of directly cooling molecules, numerous experimental efforts have been made to generate ultracold alkali molecules through associating ultracold atoms implying the Feshbach resonance \cite{Julienne06} or photoassociation \cite{Jones06} techniques. In general, the ultracold molecules created by these two approaches are distributed in various excited states. Over the last few years, a lot of theoretical and experimental investigations have demonstrated, via the stimulated Raman adiabatic passage (STIRAP) technique \cite{Shore98}, that excited ultracold molecules could be transferred to molecular ground states \cite{Winkler07}-\cite{Mackie04}.
 
 The STIRAP technique applies two specially designed sequential laser pulses to  produce a coherent population trapping (CPT) state for the purpose of
realizing a complete population transfer between two quantum states of an atom or a molecule \cite{Shore98,Bergmann03}.  In STIRAP experiments, when two pulsed lasers interact with an atom or a molecule, the corresponding Hamiltonian for the atom or molecule is time dependent. Theoretical and experimental investigations for the STIRAP have shown that it is critical to keep the atomic or molecular system  adiabatically staying in the instantaneous CPT eigenstate of the time-dependent Hamiltonian \cite{Bohm51}-\cite{Mark09}. Therefore, any dynamical deviations from the CPT state must be small enough to achieve a good STIRAP. In photoassociation of ultracold molecules combined with STIRAP, the adiabaticity analysis of the CPT state gets more complicated, because the physical process of atom-molecule transfer is intrinsically nonlinear. Two nonlinear resources exist. One is due to merging of two atoms, and the other arises from collisions between atoms and molecules. The nonlinearity could make the system unstable, such that a small deviation from the CPT state might be amplified, and consequently the transfer efficiency from atoms to molecules will be decreased \cite{Ling05}. So adiabaticity and stability are crucial for achieving an efficient molecular generation.  
Ignoring collisional interactions between atoms and molecules, adiabatic conditions for the CPT state in photoassociation process have been intensively investigated \cite{Mackie00}-\cite{Qian10}; in particular, reference \cite{Pu07} presents a general approach to treat the adiabatic condition through a linearization procedure. However, these investigations neglect the effect of nonlinear instability caused by collisional interactions between atoms and molecules \cite{Hope01}. Quite recently, references.\cite{Itin07,Itin08} extend the approach in \cite{Pu07} including dynamical instabilities, and obtain improved adiabatic conditions.

In this paper we study the dynamics of applying STIRAP to photoassociate molecules from an atomic condensate.  We emphasize the relations between dynamical instability aroused from intra-atomic collision, the adiabatic condition of a CPT state and the transfer efficiency from atoms to molecules.  We obtain an analytical formula for a CPT state including the atom-atom collisional effect and obtain the collision-induced frequency shift between two quantum states. Further, even in dynamically unstable regimes, we find that the adiabatic realization of the CPT state is available, and a very efficient atom-molecule transfer could be achieved when the single-photon detuning can compensate the collision-induced frequency shift. 

Our paper is organized as follows. In section 2, we present  a two-color $\Lambda$-type model in the mean-field treatment and study the stability and adiabaticity properties for the corresponding CPT state. In particular, we derive an analytical form for the adiabatic condition in the case of effective single-photon resonance. In section 3, we numerically test our theoretical results and discuss the experimental feasibility. Finally, a summary is presented in section 4.

\section{Theory: a nonlinear $\Lambda$-model}

\subsection{CPT State with atom-atom collision} \label{CPT distributions}

\begin{figure}[ptb]%
\centering
\includegraphics[
width=3.0in
]%
{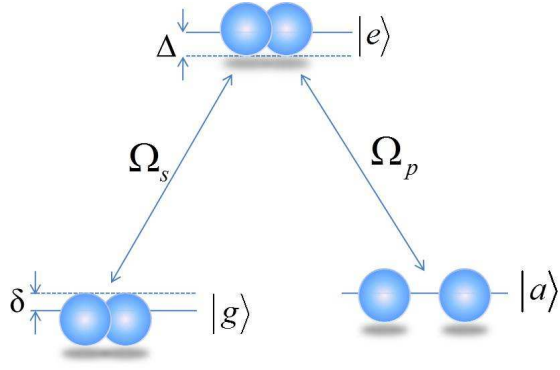}%
\caption{$\Lambda$-type configuration for an atom-molecule conversion scheme. Parameters are described in the text.}%
\label{model}%
\end{figure}

Our scheme of photoassociating molecules applies a two-color $\Lambda$-type
configuration \cite{Gaubatz88,Kuklinski89} as shown in figure \ref{model}. First, a pair of ultracold atoms initially occupied in the ground state $\left\vert a\right\rangle$ is associated with and excited to a molecular state  $\left\vert e\right\rangle $ via a pump laser with a one-photon detuning $\Delta$ from the state $\left\vert e\right\rangle $. Subsequently, another control laser with a detuning $\Delta$+$\delta$ ($\delta$, the two-photon detuning) from the molecular excited state $\left\vert e\right\rangle $ takes the excited molecules to the molecular ground state $\left\vert g\right\rangle $. The Rabi frequencies by the two lasers, respectively denoted as $\Omega_{p}$ and $\Omega_{s}$, are treated as real numbers in our following discussions.

In the interaction picture, the Hamiltonian for the atom-molecule system is given by \cite{Jing08}:
\begin{eqnarray}
\hat{H} &=&\hbar \int d\mathbf{r}\left\{ -\Delta \hat{\psi}_{e}^{\dagger
}\left( \mathbf{r}\right) \hat{\psi}_{e}\left( \mathbf{r}\right) -\delta 
\hat{\psi}_{g}^{\dagger }\left( \mathbf{r}\right) \hat{\psi}_{g}\left( 
\mathbf{r}\right) -\frac{1}{2}\left( \Omega _{p}^{\prime}\hat{\psi}_{e}^{\dagger
}\left( \mathbf{r}\right) \hat{\psi}_{a}\left( \mathbf{r}\right) \hat{\psi}%
_{a}\left( \mathbf{r}\right) \right. \right.  \nonumber   \\
&&\left. \left. +\Omega _{s}\hat{\psi}_{g}^{\dagger }\left( \mathbf{r}%
\right) \hat{\psi}_{e}\left( \mathbf{r}\right) +H.c.\right) +U_{aa}^{\prime}\hat{\psi}%
_{a}^{\dagger }\left( \mathbf{r}\right) \hat{\psi}_{a}^{\dagger }\left( 
\mathbf{r}\right) \hat{\psi}_{a}\left( \mathbf{r}\right) \hat{\psi}%
_{a}\left( \mathbf{r}\right) \right\} \label{Hamn}
\end{eqnarray}
where $\hat{\psi}_{i}\left(  \hat{\psi}_{i}^{\dagger}\right) $ are the
annihilation(creation) field operators for the state $\left\vert
i\right\rangle $ $(i=a,e,g)$. In equation (\ref{Hamn}), $U_{aa}^{\prime}=4\pi\hbar a_{s}/m$ with $s$-wave scattering length $ a_{s}$ is the pseudo-potential for the atom-atom interaction. Atom-molecule and molecule-molecule interactions are ignored  for the purpose of extracting physics induced by two-body $s$-wave collisions. Actually, our later numerical simulations show that by adding the neglected collisional effects, our main results will not change.  

In our analysis, we are concerned with a spatially uniform system at zero-temperature, and thus the kinetic and trapping potential terms within equation (\ref{Hamn}) have been safely removed. Using the mean-field method that replace
$\hat{\psi}_{i}^{(\dagger)}$ with $\sqrt{n}\psi
_{i}^{\left( \ast \right) }$, where $n$ is the density of the total particle number, from equation (\ref{Hamn}), we obtain a set of coupled Heisenberg motional equations for $\sqrt{n}\psi_{i}^{\left( \ast \right) }$, given by
\begin{eqnarray}
i\dot{\psi}_{a}  &  =-\Omega_{p}\psi_{a}^{\ast}\psi_{e}+2U_{aa}\left\vert
\psi_{a}\right\vert ^{2}\psi_{a}\label{dy1}\\
i\dot{\psi}_{e}  &  =-\left(  \Delta+i\gamma\right)  \psi_{e}-\frac{\Omega
_{p}}{2}\psi_{a}^{2}-\frac{\Omega_{s}}{2}\psi_{g}\label{dy2}\\
i\dot{\psi}_{g}  &  =-\delta\psi_{g}-\frac{\Omega_{s}}{2}\psi_{e}\label{dy3}%
\end{eqnarray}
where $\gamma$ is introduced phenomenologically for describing
the spontaneous emission of the excited state $\left\vert e\right\rangle $,
 $\Omega_{p}= \sqrt{n}\Omega_{p}^{\prime}$ and $U_{aa}=nU_{aa}^{\prime}$.

Without an intra-atomic collisional term in equation (\ref{dy1}), a CPT state has already been obtained \cite{Pu07}. Now, including the collision, how the CPT state is modified is worthy of investigating \cite{Ling04}. To obtain a CPT state including intra-atomic collisional effect, we assume $\psi_{a}(t)=\phi_{a}e^{-i\mu t}$, $\psi_{e}=0$, $\psi_{g}(t)=\phi_{g}e^{-2i\mu t}$, where $\mu$ is the chemical potential and $\phi_{i}(i=a,e,g)$ are for population distributions in the CPT state. We also neglect the spontaneous loss $\gamma$ and treat all the other parameters as time-independent. Inserting the ansatz for the CPT state into equations (\ref{dy1})-(\ref{dy3}) and considering the total particle number normalization,
 $\phi_{a}
^{2}+2\phi_{g} ^{2}  =1$, we obtain the population distributions
given by a new vector $\mathbf{\lambda}=[\phi_{a},\phi_{e},\phi_{g}]^{T}$, where
\begin{equation}
\phi_{a}=\sqrt{\frac{2}{\sqrt{1+8\chi^{2}}+1}},\phi_{e}=0,\phi_{g}=\frac{-2\chi}%
{1+\sqrt{1+8\chi^{2}}} \label{dark_popu}%
\end{equation}
under the condition of an effective two-photon resonance
$\delta_{eff}\equiv\delta+4U_{aa}\phi_{a}^{2}=0$ with $\chi=\Omega_{p}/\Omega_{s}$. Here, $4U_{aa}\phi_{a}^{2}$ is the collision-induced frequency shift. The chemical potential corresponding to equation (\ref{dark_popu}) is $\mu=2U_{aa}\phi_{a}^{2}$.

Equation(\ref{dark_popu}) shows that all the atoms $\phi_{a}^{2}$ could be converted into ground-state molecules $\phi_{g}^{2}$ under the two-photon resonance when $\chi$ approaches to $\infty$. More interestingly,  equation (\ref{dark_popu}) indicates that when $\delta=-4U_{aa}\phi_{a}^{2}$ is satisfied, the CPT state does not rely on the $U_{aa}$ coefficient. In contrast, previous works on other models for photoassociating molecules shows that the atom-atom collision could easily lead to the deviation of the system from a CPT state, and thus the final produce of molecules in the ground state could  be low \cite{Kokklemans01}.

\subsection{Analysis on stability and adiabaticity of the CPT state with atom-atom collision }

 In section \ref{CPT distributions}, we have derived a CPT state with atom-atom collision. However, this collision could lead to dynamical instability of the atom-molecule coupled system. In this situation, whether the CPT state could be adiabatically followed is worthy of further exploring. In this section, we start to investigate an adiabatic theory for the CPT state.    

First, we add small perturbations $\delta\psi_{i}\left(  t\right)  $ to the CPT state distributions $\phi_{i}\left(
t\right)$, i.e., 
\begin{eqnarray}
\psi_{a}\left(  t\right)   &  =\left(  \phi_{a}\left(  t\right)  +\delta
\psi_{a}\left(  t\right)  \right)  e^{-i\mu t}\label{linearized1}\\
\psi_{e}\left(  t\right)   &  =\delta\psi_{e}\left(  t\right)  e^{-2i\mu t}\label{linearized2}\\
\psi_{g}\left(  t\right)   &  =\left(  \phi_{g}\left(  t\right)  +\delta
\psi_{g}\left(  t\right)  \right)  e^{-2i\mu t}\label{linearized3}%
\end{eqnarray}
and study the time evolution of the perturbations. Inserting equations (\ref{linearized1})-(\ref{linearized3}) into equations (\ref{dy1})-(\ref{dy3}) and taking into account the time dependence of
the Rabi frequencies $\Omega_{p,s}\left(  t\right)  $ whose time dependence is ignored in section \ref{CPT distributions}, we obtain a set of linearized coupled equations for the
vector $\mathbf{\delta\psi}\left(  t\right)  =\left[  \delta\psi_{a}%
,\delta\psi_{e},\delta\psi_{g}\right]  ^{T}$ with its conjugate component
$\mathbf{\delta\psi}^{\ast}\left(  t\right)  $:%
\begin{equation}
i\mathbf{\delta\dot{\psi}}=\mathbf{A}(t)\mathbf{\delta\psi}+\mathbf{B}%
(t)\mathbf{\delta\psi}^{\ast}-i\gamma\mathbf{F\delta\psi}-i\mathbf{\dot
{\lambda}}(t) \label{psi_vec_eq}.%
\end{equation}
In equation (\ref{psi_vec_eq}), the coefficient matrices $\mathbf{A}$ and $\mathbf{B}$ are given by %
\[
\mathbf{A}=\left[
\begin{array}
[c]{ccc}%
2U_{aa}\phi_{a}^{2} & -\Omega_{p}\phi_{a} & 0\\
-\Omega_{p}\phi_{a} & -\Delta_{u} & -\frac{\Omega_{s}}{2}\\
0 & -\frac{\Omega_{s}}{2} & 0
\end{array}
\right]  ,
\mathbf{B}=\left[
\begin{array}
[c]{ccc}%
2U_{aa}\phi_{a}^{2} & 0 & 0\\
0 & 0 & 0\\
0 & 0 & 0
\end{array}
\right]
\]
with $\Delta_{u}=\Delta+4U_{aa}\phi_{a}^{2}$, $\mathbf{\dot{\lambda}%
}(t)=[\dot{\phi}_{a}(t),0,\dot{\phi}_{g}(t)]^{T}$ is a driving source term, and 
$\mathbf{F}$ is a 3$\times$3 matrix associated with the loss of the excited
state denoted by the only nonzero element $F_{22}=1$. For the convenience of the following analysis, we introduce a vector $\Xi(t)=[\mathbf{\delta\psi}(t),\mathbf{\delta\psi
}^{\ast}(t)]^{T}$, and rewrite the motional equation (\ref{psi_vec_eq}) as,%
\begin{equation}
\dot{\Xi}(t)=-i\mathbf{M}^{T}(t)\Xi(t)-\gamma\mathbf{D}\Xi(t)-\mathbf{\dot
{\Lambda}}\left(  t\right)  \label{chi_eq}%
\end{equation}
where
\[
\mathbf{M}\left(  t\right)  =%
\left [
\begin{array}{cc}
\mathbf{A}(t) & -\mathbf{B}(t)\\
\mathbf{B}(t) & -\mathbf{A}(t)
\end{array}
\right ]
,\mathbf{D}=%
\left [
\begin{array} {cc}
\mathbf{F} & 0\\
0 & \mathbf{F}%
\end{array}
\right ]
, \mathbf{\dot{\Lambda}}\left(  t\right)  =\left[
\begin{array}
[c]{c}%
\mathbf{\dot{\lambda}}(t)\\
\mathbf{\dot{\lambda}}(t)
\end{array}
\right].
\]

Second, we further write the characteristic equation for the matrix
$\mathbf{M}\left(  t\right)  $ as $\mathbf{M}(t)\mathbf{w}_{i}%
(t)=\omega_{i}(t)\mathbf{w}_{i}(t)$, where $\omega_{i}(t)$ is the
eigenvalue and $\mathbf{w}_{i}(t)=[\mathbf{u}_{i}(t),\mathbf{v}_{i}(t)]^{T}$
the $i$th eigenvector. Vectors $\mathbf{u}_{i}(t)$ and $\mathbf{v}%
_{i}(t)$ containing the familiar Bogoliubov $u$-$v$ parameters should fulfill
the normalized condition of $\sum_{j=a,e,g}\left(  u_{ij}^{2}\left(  t\right)
-v_{ij}^{2}\left(  t\right)  \right)  =1$, where vectors
$\mathbf{u}_{i}(t)=\left[  u_{ia}(t),u_{ie}(t),u_{ig}(t)\right]  ^{T}$ and
$\mathbf{v}_{i}(t)=\left[  v_{ia}(t),v_{ie}(t),v_{ig}(t)\right]  ^{T}$.
By solving the corresponding secular equation of matrix $\mathbf{M}\left(
t\right)  $, we obtain eigenvalues $\omega_{i}(t)$, which are a pair of zero frequency mode $\omega_{0}=0$ and two pairs of
nonzero excited modes $\pm\omega_{1,2}\left(  t\right)  $ given by%
\begin{equation}
\omega_{1,2}(t)=\sqrt{a_{1}\pm\sqrt{a_{1}^{2}-a_{2}}} \label{eigenvalue}%
\end{equation}
with $a_{1}=\Omega_{s}\Omega_{eff}/4+\Delta_{u}^{2}/2$, $a_{2}=\Omega_{s}%
^{2}\Omega_{eff}^{2}/16+4U_{aa}\Omega_{p}^{2}\phi_{a}^{4}\Delta_{u}$,
$\Omega_{eff}=\sqrt{\Omega_{s}^{2}+8\Omega_{p}^{2}}$. We can prove that without atom-atom collision($U_{aa}=0$), $\omega_{1,2}(t)$ are real and the CPT state is dynamically stable. However, when the atom-atom collision is present, in the spaces of $a_{2}<0$ or $a_{2}>a_{1}^{2}$, $\omega_{1,2}(t)$ are not real and thus the CPT state solution becomes unstable.

\begin{figure}[ptb]%
\centering
\includegraphics[
height=2.5in,
width=3.0in
]%
{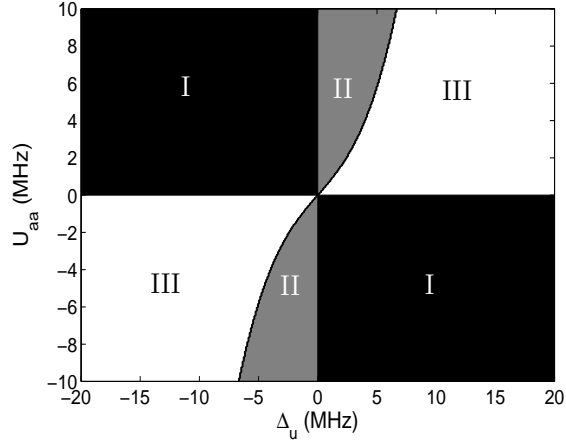}%
\caption{Stability diagram with ($U_{a}$, $\Delta_{u}$) space.
Parameters used in our calculations are defined by $\Omega_{p,s}^{0}=10$MHz, $t_{p}%
=19\mu$s, $t_{s}=11\mu$s, $\tau$=4$\mu$s.}%
\label{stability_dia}%
\end{figure}

In experiments, $U_{aa}$ and
$\Delta_{u}$ are two controllable parameters. Thereby, we study the dynamical stability for the CPT in ($U_{aa}$,
$\Delta_{u}$) space. To get an intuitive picture for the stability, we plot a stability diagram in figure \ref{stability_dia}, where a pair of Gaussian pulses is adopted with the pulse width $\tau$:
\begin{equation}
\Omega_{p,s}=\Omega_{p,s}^{0}\exp\left[ -\frac{\left ( t-t_{p,s}\right)
^{2}}{\tau^{2}}\right] \label{Gaussian}%
\end{equation}
where $t_{p,s}$ and $\Omega_{p,s}^{0}$ are the central positions and peak strengths, respectively. In our calculations, a pair of fixed laser pulses is applied:
$\Omega_{p,s}^{0}=10^{7}$s$^{-1}$, $t_{p}%
=19\mu$s, $t_{s}=11\mu$s, $\tau$=4$\mu$s. Figure \ref{stability_dia} shows unstable regions, respectively, labeled by \textrm{I}($a_{2}<0$), \textrm{II}($a_{2}>a_{1}^{2}$), and stable regions labeled by \textrm{III}. Strict mathematical results for these regions are listed below:


Region \textrm{I} is for $a_{2}<0$, given by
\[
 U_{aa}>0,   \Delta _{u} <-\frac{\Omega _{eff}^{2}}{64U_{aa}\chi ^{2}\phi _{a}^{4}};
 U_{aa}<0,   \Delta _{u} >-\frac{\Omega _{eff}^{2}}{64U_{aa}\chi ^{2}\phi _{a}^{4}}.%
\]

Region \textrm{II} is for $a_{2}>a_{1}^{2}$, given by%
\[
\Delta _{u}\in \left( \min \left( x_{0},0\right) ,\max \left( x_{0},0\right)
\right) 
\]%
where $x_{0}$ is the real root derived from a cubic equation $x^{3}+\Omega
_{s}\Omega _{eff}x-16U_{aa}\Omega _{p}^{2}\phi _{a}^{4}=0$.

Region \textrm{III} is for dynamical stable regimes.

 After deciding the stability properties, we proceed to study the adiabaticity of the CPT state. We turn our attention to eigenstates of the matrix $\mathbf{M}(t)$.
 The eigenstate with $\omega_{0}=0$ corresponding to the Goldstone mode
takes a non-normalized form of%
\begin{equation}
\mathbf{w}_{0}(t)=\left(  \frac{\Omega_{s}}{2},0,-\Omega_{p}\phi_{a}%
,\frac{\Omega_{s}}{2},0,-\Omega_{p}\phi_{a}\right)  ^{T}\equiv\mathbf{P}
\label{(eigenvector_eq)}%
\end{equation}
 
Since the Goldstone mode is degenerate, we need to introduce a new vector $\mathbf{Q}$ complimentary to $\mathbf{P}$ given by\cite{Ling07}%

\begin{equation}
\mathbf{MQ}=\frac{\mathbf{P}}{\nu} \label{MQ_eq}%
\end{equation}
where $\nu$\ is determined from a normalization condition
\begin{equation}
\mathbf{Q}^{\dag}\eta_{+}\mathbf{P}=1 \label{normalized}%
\end{equation}
and $\eta_{+}$ is given in equation (\ref{eta_pm}).
By solving equations (\ref{MQ_eq}) and (\ref{normalized}) simultaneously, we find
$\mathbf{Q}$ has the following structure:%
\begin{equation}
\mathbf{Q}=(q_{a},q_{e},q_{g},-q_{a},-q_{e},-q_{g})^{T} \label{Q_vector}%
\end{equation}
where elements $q_{i}$ are%
\[
q_{a}=\frac{\Omega _{eff}}{8vU_{aa}\phi _{a}^{2}},q_{e}=\frac{2\chi \phi _{a}%
}{v},q_{g}=-\frac{16U_{aa}\phi _{a}^{2}\chi \Delta _{u}+\Omega _{p}\Omega
_{eff}}{4vU_{aa}\phi _{a}\Omega _{s}}
\]
and the coefficient $v$ is $v=\left(  64U_{aa}\phi_{a}^{4}\chi^{2}\Delta
_{u}+\Omega_{eff}^{2}\right)  /8U_{a}\phi_{a}^{2}$.
All of the base vectors including $\mathbf{P}$, $\mathbf{Q}$, $\mathbf{w}_{j}$ corresponding to
$\pm\omega_{j}(j=1,2)$ create a complete 6$\times$6 space in the sense that they obey the
following biorthonormality relations:%
\begin{equation}
\mathbf{w}_{i}^{\dagger}\eta_{+}\mathbf{w}_{j}    =\delta_{ij},\mathbf{w}%
_{i}^{\dagger}\eta_{-}\mathbf{w}_{j}=0\nonumber
\end{equation}
\begin{equation}
\mathbf{P}^{\dagger}\eta_{\pm}\mathbf{w}_{j}    =0,\mathbf{Q}^{\dagger}%
\eta_{\pm}\mathbf{w}_{j}=0\label{biorth}
\end{equation}
where $\eta_{+}$ and $\eta_{-}$ are two metric matrices defined as%
\begin{equation}
\eta_{+}=\left[
\begin{array}
[c]{cc}%
\mathbf{I} & 0\\
0 & -\mathbf{I}%
\end{array}
\right]  ,\eta_{-}=\left[
\begin{array}
[c]{cc}%
0 & \mathbf{I}\\
-\mathbf{I} & 0
\end{array}
\right]  \label{eta_pm}%
\end{equation}
and $\mathbf{I}$ being a 3$\times$3 unit matrix.
Equipped with this totally complete space, we are able to expand $\Xi(t)$\ in
the form of:%
\begin{equation}
\Xi(t)=c_{p}\eta_{+}\mathbf{Q}+c_{q}\eta_{+}\mathbf{P}+\sum_{i=1,2}(c_{i}%
\eta_{+}\mathbf{w}_{i}-c_{i}^{\ast}\eta_{-}\mathbf{w}_{i}^{\ast})
\label{superp_chi}%
\end{equation}
Inserting equation (\ref{superp_chi}) into equation (\ref{chi_eq}) and using biorthonormality relations equations (17) and (18), one could obtain the dynamic equations for
$c_{p,q}$:
\begin{equation}
\frac{dc_{p}}{dt}=0,\frac{dc_{q}}{dt}=-i\frac{c_{p}}{v} \label{goldstonen}%
\end{equation}
In deriving equation (\ref{goldstonen}), we have assumed that the vectors $\mathbf{P}$ and
$\mathbf{Q}$ are time independent, so that the source terms $\mathbf{P}^{\dagger
}\left(  \mathbf{Q}^{\dagger}\right)  \mathbf{\dot{\Lambda}}(t)$ will
vanish. Thereby, $\dot{c}_{p}$ and $\dot{c}_{q}$
would be decoupled from the motional equation for the vector $\Xi(t)$. Apparently, here we only need to focus on the values $c_{1,2}(t)$ and $c_{1,2}^{\ast}(t)$ for the purpose of adiabatic condition. The dynamical
behavior for $c_{i}(t)\left(  i=1,2\right)  $ is governed by%
\begin{equation}
\frac{dc_{i}}{dt}+i\omega_{i}^{\ast}c_{i}+\gamma\mathbf{w}_{i}^{\dagger
}\mathbf{D\chi}(t)=-\mathbf{w}_{i}^{\dagger}\mathbf{\dot{\Lambda}}\left(
t\right)  \label{cj_dys}%
\end{equation}
In the adiabatic limit, $\dot{\Omega}_{p,d}\left(  t\right)  \rightarrow0$,
'projections' $c_{i}$ can be treated as time-independent values since $\omega_{i}$\ and $\gamma$ are usually
sufficiently large. Now, we estimate $c_{i}$\ up to the first order
of $\dot{\Omega}_{p,d}(t)$,  and obtain  the following set of linearly coupled
equations:%
\begin{equation}
\begin{array}[c]{l} %
  \left(
\begin{array}
[c]{cccc}%
i\omega_{1}^{\ast}+\gamma g_{11} & \gamma g_{12} & {0} & -\gamma
f_{12}\\
\gamma g_{12}^{\ast} & i\omega_{2}^{\ast}+\gamma g_{22} & \gamma f_{12} & 0\\
0 & -\gamma f_{12}^{\ast} & \gamma g_{11}^{\ast}-i\omega_{1} & \gamma
g_{12}^{\ast}\\
\gamma f_{12}^{\ast} & 0 & \gamma g_{12} & \gamma g_{22}^{\ast}-i\omega_{2}%
\end{array}
\right)  \left(
\begin{array}
[c]{c}%
c_{1}\\
c_{2}\\
c_{1}^{\ast}\\
c_{2}^{\ast}
 \end{array}
 \right)
\\  \nonumber
 =\left(
\begin{array}
[c]{c}%
-\mathbf{w}_{1}^{\dag}\mathbf{\dot{\Lambda}}(t)\\
-\mathbf{w}_{2}^{\dag}\mathbf{\dot{\Lambda}}(t)\\
-\mathbf{w}_{1}^{T}\mathbf{\dot{\Lambda}}(t)\\
-\mathbf{w}_{2}^{T}\mathbf{\dot{\Lambda}}(t)
\end{array}
  \right)
  \end{array}
    \label{coupled_excit}%
\end{equation}
where coefficients $g_{ij}$ and $f_{ij}$ are denoted by %
$g_{ij}=u_{ie}^{\ast}u_{je}-v_{ie}^{\ast}v_{je},f_{ij}=u_{ie}^{\ast}%
v_{je}^{\ast}-v_{ie}^{\ast}u_{je}^{\ast} \label{gf_eq} $
and%
\begin{equation}
\mathbf{w}_{j}^{\dag}\mathbf{\dot{\Lambda}}\left(  t\right)  =%
\frac{\phi_{g}\left(  \dot{\Omega}_{p}-\chi\dot{\Omega}_{s}\right)
\left\{  2\chi\left(  u_{ja}^{\ast}+v_{ja}^{\ast}\right)  \phi
_{a}+\left( u_{jg}^{\ast}+v_{jg}^{\ast}\right)  \right\}
}{\chi \Omega_{eff}} \label{source_term}%
\end{equation}

As for $\mathbf{w}_{j}^{T}\mathbf{\dot{\Lambda}}\left(  t\right)(j=1,2)  $,
$u^{\ast}\left(  v^{\ast}\right)  _{ja\left(  g\right)  }$ is replaced by
$u\left(  v\right)  _{ja\left(  g\right)  }$ in equation (\ref{source_term}).
Compared with the results of equation (42) in \cite{Ling07},  equation (\ref{coupled_excit}) has been extended to include unstable parameter regions.  

In \cite{Pu07,Ling07}, an adiabatic parameter 
\begin{equation}
r=\frac{\sqrt{\left\vert c_{1}\right\vert ^{2}+\left\vert c_{2}\right\vert
^{2}}}{2}\ll1 \label{r_parameters}%
\end{equation}
is introduced to measure the quality of adiabaticity. In general, $r\ll1$ is required to keep good adiabaticity of the CPT state. After solving equation (\ref{coupled_excit}), we could calculate the adiabatic parameter $r$. However,  the general analytical formula is very complicated for analysis. In the following, we calculate the adiabatic parameter  in a special case of $\Delta_{u}=0$.

\subsection{Adiabaticity in the case of $\Delta_{u}=0$} \label{adia}

In the derivation of adiabatic parameter $r$, we are pleased to
see if $\Delta_{u}$ is tuned to 0, i.e. on effective one-photon resonance,
equation (\ref{chi_eq}) can be further simplified to
\begin{equation}
\dot{\Xi}^{\prime}\left( t\right)=-i\mathbf{M^{\prime}}\left(  t\right)\Xi^{\prime
}\left(  t\right)  -\gamma\mathbf{F}\Xi^{\prime
}\left(  t\right)  -2\dot{\lambda}\left(  t\right)
\label{Delta_u_eqs}%
\end{equation}
where the vector $\Xi^{\prime}\left(  t\right)  $ is described as $\Xi%
^{\prime}=\left[  \delta\psi_{a}^{+},\delta\psi_{e}^{-},\delta\psi_{g}%
^{+}\right]  ^{T}$ with elements $\delta\psi_{j}^{\pm}=\delta\psi_{j}\pm
\delta\psi_{j}^{\ast}$($j=a,e,g$). In this case, we find the coupling coefficient matrix $\mathbf{M}%
^{\prime}\left(  t\right)  $ turns out to be%
\begin{equation}
\mathbf{M}^{\prime}\left(  t\right)  =\left(
\begin{array}
[c]{ccc}%
0 & -\Omega_{p}\phi_{a} & 0\\
-\Omega_{p}\phi_{a} & 0 & -\frac{\Omega_{s}}{2}\\
0 & -\frac{\Omega_{s}}{2} & 0
\end{array}
\right)  \label{matrix_p}%
\end{equation}
Surprisingly, in contrast to $\mathbf{M}\left(  t\right)  $ in equation (\ref{chi_eq}), now $\mathbf{M}%
^{\prime}\left(  t\right)  $ becomes 3$\times$3 dimensional and symmetrical,
and its eigenvalues and eigenvectors are all analytically solvable, which are%
\begin{equation}
\omega_{0}^{\prime}=0,\omega_{\pm}^{\prime}=\pm\frac{1}{2}\sqrt{\Omega
_{s}\Omega_{eff}} \label{eigenvalues}%
\end{equation}
and
\begin{eqnarray}
\mathbf{w}_{0}^{\prime}  &  =\frac{1}{\sqrt{\Omega_{s}\Omega_{eff}}}\left[
-\Omega_{s},0,2\phi_{a}\Omega_{p}\right]  ^{T},\label{eigenvectors1} \\
\mathbf{w}_{\pm}^{\prime}  &  =\frac{1}{\sqrt{2\Omega_{s}\Omega_{eff}}}\left[
2\phi_{a}\Omega_{p},\pm\sqrt{\Omega_{s}\Omega_{eff}},\Omega_{s}\right]  ^{T}.\label{eigenvectors2}%
\end{eqnarray}
More remarkable, the eigenvectors become complete and orthogonal.  As a result,
any bare vector $\mathbf{\delta\psi}\left(  t\right)  $ in
equation (\ref{psi_vec_eq}) can be directly expanded by $\mathbf{w}_{j}^{\prime}$,
i.e. $\mathbf{\delta\psi}\left(  t\right)  =\sum_{i=0,\pm}c_{j}\mathbf{w}%
_{j}^{\prime}$, leading to a new set of linearly coupled equations,%
\begin{equation}
i\left(
\begin{array}
[c]{cc}%
\mathbf{C} & \mathbf{D}\\
-\mathbf{D} & -\mathbf{C}%
\end{array}
\right)  \left(
\begin{array}
[c]{c}%
\mathbf{\dot{c}}\\
\mathbf{\dot{c}}^{\ast}%
\end{array}
\right)  +\left(
\begin{array}
[c]{cc}%
\mathbf{\Gamma} & 0\\
0 & \mathbf{\Gamma}%
\end{array}
\right)  \left(
\begin{array}
[c]{c}%
\mathbf{\dot{c}}\\
\mathbf{\dot{c}}^{\ast}%
\end{array}
\right)  =-\left(
\begin{array}
[c]{c}%
\mathbf{\Pi}\\
\mathbf{\Pi}%
\end{array}
\right).  \label{n_coupled_eq}%
\end{equation}
Here vectors $\mathbf{c=}\left[  c_{+},c_{-}\right]  ^{T}$, $\mathbf{\Pi
}=\left[  \mathbf{w}_{+}^{\prime T}\mathbf{\dot{\lambda},w}_{-}^{\prime
T}\mathbf{\dot{\lambda}}\right]  ^{T}$ and other coupled coefficient matrices
are%
\[
\mathbf{C=}\left(
\begin{array}
[c]{cc}%
\kappa_{-} & \kappa_{0}\\
\kappa_{0} & \kappa_{+}%
\end{array}
\right)  ,\mathbf{D=}\left(
\begin{array}
[c]{cc}%
\kappa_{0} & \kappa_{0}\\
\kappa_{0} & \kappa_{0}
\end{array}
\right)  ,\mathbf{\Gamma}=\left(
\begin{array}
[c]{cc}%
\frac{\gamma}{2} & -\frac{\gamma}{2}\\
-\frac{\gamma}{2} & \frac{\gamma}{2}%
\end{array}
\right)
\]
with the corresponding elements
\[
\kappa_{\pm}=\frac{8U_{aa}\Omega_{p}^{2}\phi_{a}^{4}\pm\left(  \Omega
_{s}\Omega_{eff}\right)  ^{3/2}}{2\Omega_{s}\Omega_{eff}},
\kappa_{0}=\frac{1}{2}\left(\kappa_{+}+\kappa_{-}\right).%
\]

In obtaining equation (\ref{n_coupled_eq}), we consider $c_{j}\rightarrow0$ for
the adiabatic limit. In addition, the zero modes $c_{0}$ and
$c_{0}^{\ast}$ are left decoupled due to the weak coupling strengths and
the eliminated source terms. Based on the above analysis, the adiabatic parameter $r$
can be solved analytically%
\begin{equation}
r=\frac{\sqrt{\left( 4\gamma ^{2}+\Omega _{s}\Omega _{eff}\right) \left(
1+\eta \right) }\left\vert \dot{\chi}\right\vert }{\Omega _{s}^{1/2}\Omega
_{eff}^{3/2}\left( 1+\sqrt{1+8\chi ^{2}}\right) /2}  \label{ann}
\end{equation}
with $\eta =\left( 1-\frac{1}{\sqrt{1+8\chi
^{2}}}\right) ^{4}\frac{16U_{aa}^{2}\gamma ^{2}}{\Omega _{p}^{4}}$.
 
Equation(\ref{ann}) agrees with equation (11) in \cite{Pu07} in the case of $\gamma=U_{aa}=0$.  In equation (\ref{ann}), only the parameter $\eta$ is related to the atom-atom collisional strength. Evidently, when $\eta\ll 1$ which could be met by using strong pump laser such that $ 16U_{aa}^{2}\gamma ^{2}/\Omega _{p}^{4} \ll 1$, the adiabatic parameter $r$ is nearly immune to atom-atom collisional effect.

\section{Numerical Results}

\subsection{Stability and Adiabaticity} \label{SA}

\begin{figure}[ptb]%
\centering
\includegraphics[
height=2.687375in,
width=3.5046in
]%
{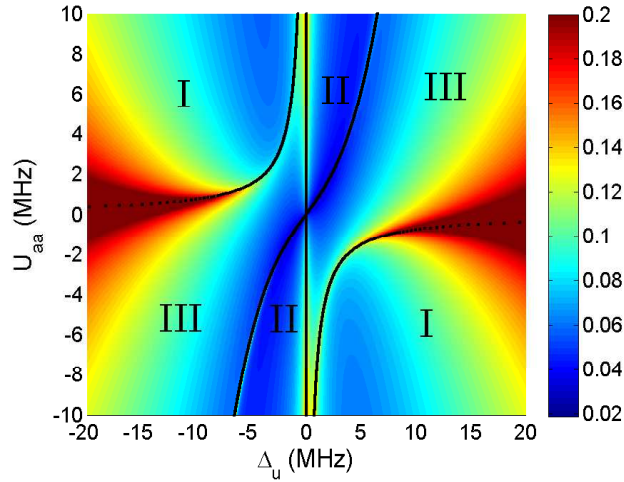}%
\caption{(Color online)Adiabaticity parameter $r$-value as a function of $U_{aa}$ and
$\Delta_{u}$ at $t=t_{sp}$. Laser
pulses are the same as used in figure \ref{stability_dia}, the spontaneous loss is
$\gamma=1.0$MHz. Black curves describe the corresponding instability
boundaries at $t=t_{sp}$. }%
\label{adiabaticity_pic}%
\end{figure}

In this subsection, we numerically study the relation of the stability and adiabaticity for the CPT state with atom-atom collision. 
In figure \ref{adiabaticity_pic}, we plot the distributions of the adiabatic parameter $r$ defined in equation (\ref{r_parameters}) in $(U_{aa},\Delta_{u})$ parameter space at time $t=t_{sp}\equiv(t_{p}+t_{s})/2$. 
The unstable and stable regions, whose symbols \textrm{I}, \textrm{II} and \textrm{III} are the same as those used in figure \ref{stability_dia}, are also shown in figure \ref{adiabaticity_pic}. Black curves stand for the boundary of unstable regions at $t=t_{sp}$. It is interesting to see that in the region near the boundary of the stable region \textrm{III} and unstable region \textrm{II}, $r$ is very small ($<$0.1) no matter how strong the atom-atom collision $U_{aa}$ is. And, $r$ at the right corner of the upper unstable region \textrm{I} is even lower than those close to this corner but in the stable region \textrm{III}. So, in our scheme, good adiabaticity could even be achieved in the dynamical unstable regions.

\begin{figure}[ptb]%
\centering
\includegraphics[
height=2.6in
]%
{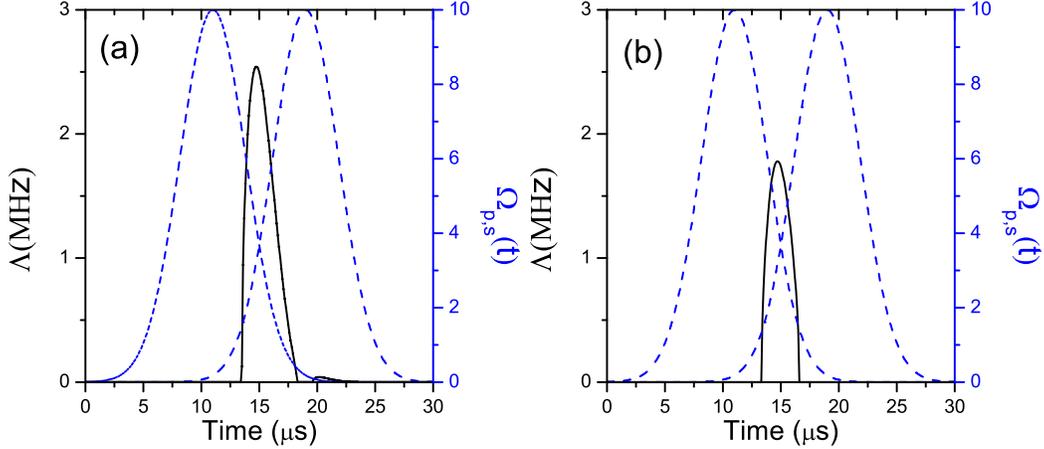}%
\caption{(a) and (b) respectively correspond to the unstable region \textrm{I} with parameters $U_{aa}=8$MHz, $\Delta_{u}=-3$MHz and unstable region \textrm{II} with parameters $U_{aa}=8$MHz, $\Delta_{u}=3$MHz, showing the unstable growth rate $\Lambda$ and Rabi frequencies $\Omega_{p,s}$ of pump and control lasers in the time domain.  }%
\label{omega_com_pic}%
\end{figure}

It is widely accepted that the $r$ values in unstable regions are in general larger than those in stable regions and hence good adiabaticity could be hardly achieved in the dynamical unstable parameter regions \cite{Itin07}. So, our result is counterintuitive. To understand this result, we introduce $\Lambda =|$Im$(\omega_{1})|$($\omega_{1} $ and $\omega_{2}$ are conjugate with each other) in which Im denotes the imaginary part of a complex number. $\Lambda$ measures unstable growth rate. We plot the time evolution of  $\Lambda$ in figure \ref{omega_com_pic}, respectively, in unstable region \textrm{I} with parameters $U_{aa}=8$MHz, $\Delta_{u}=-3$MHz(figure \ref{omega_com_pic}(a)), and in unstable region \textrm{II} with parameters $U_{aa}=8$MHz, $\Delta_{u}=3$MHz(figure \ref{omega_com_pic}(b)). 
Figure \ref{omega_com_pic} shows that the unstable growth rates in both regions have a Gaussian-like shape with a very small time width. The peak value of the unstable growth rate in the unstable region \textrm{II} is smaller than that in the unstable region \textrm{I}, so the adiabatic parameter   
in the unstable region \textrm{II} is smaller than that in the unstable region \textrm{I}.  In figure \ref{omega_com_pic}, we also plot the time evolutions of the pump and control laser intensities $\Omega_{p,s}$. When two lasers overlap in time domain, the unstable growth rate is nonzero. Thus, if the overlapping time of the two lasers is short enough, the unstable growth of the deviation from the CPT state could be negligible and then a good adiabaticity could be maintained in this situation.  

 \subsection{Transfer Efficiency} \label{TE}

\begin{figure}[ptb]%
\centering
\includegraphics[
height=2.589in,
width=3.377125in
]%
{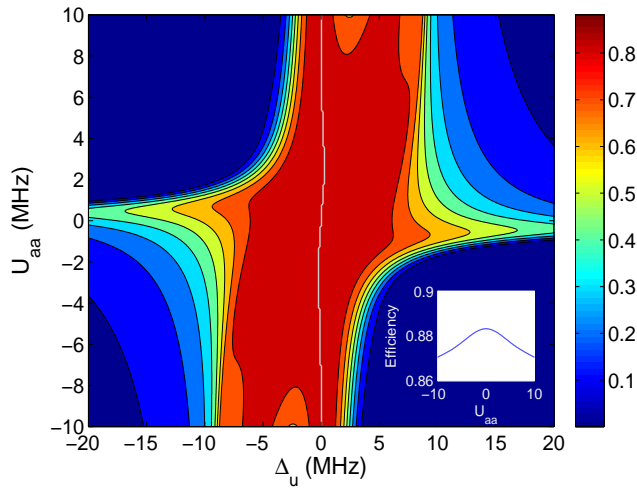}%
\caption{(Color online)Atom-molecule transfer efficiency in $\left(  U_{aa},\Delta
_{u}\right)  $ space. The central white
curve represents the maximum efficiencies with different $U_{aa}$ values.
Inset: Mapping of  the white curve as a function of $U_{aa}$,
displaying an axis-symmetrical pattern with $U_{aa}=0$.
Parameters $\Omega_{p,s}^{0}$, $t_{p,s}$ and $\tau$ used are the same as in figure \ref{stability_dia}.}%
\label{population_dy}%
\end{figure}

In this section, we further investigate the relations of the atom-molecule transfer efficiency to the stability and adiabaticity of a CPT state. The Heisenberg mean-field equations (\ref{dy1})-(\ref{dy3}) are numerically solved using the same parameters ($\Omega_{p,s}^{0},\tau,t_{p},t_{s}$) as those used in figure \ref{stability_dia}. Figure  \ref{population_dy} shows the final atom-molecule transfer efficiency in ($U_{aa},\Delta_{u}$) parameter space, which are obtained when intensities of two pulsed optical fields are essentially zero after a long time. Comparing figure \ref{population_dy} with figure \ref{adiabaticity_pic}, we find that in unstable region \textrm{I}, the transfer efficiency is always low ($<$10$\%$), and the high transfer efficiency ($>$80$\%$) is achieved in the region with very good adiabaticity (the unstable region \textrm{II} and the region that is part of the stable region \textrm{III} but is also close to region \textrm{II}). 
In figure \ref{population_dy}, a nearly vertical white line close to $\Delta
_{u}=0$ indicates the maximum transfer efficiencies according to different $U_{aa}$ values. The details of the transfer efficiency along this line versus $U_{aa}$ are plotted in the inset. Obviously, by employing the effective one-photon resonance condition, the atom-molecule transfer efficiency decreases very slowly, just from 88$\%$ at $U_{aa}=0$ to 87$\%$ at $U_{aa}=10.0$MHz. That is quite a small decrease compared with a large change of
$U_{aa}$ values. Actually, this is due to insensitivity of the adiabaticity to the collisional strength between atoms, as discussed at the end of section 2. 

\begin{figure}[ptb]%
\centering
\includegraphics[
width=4.2in
]%
{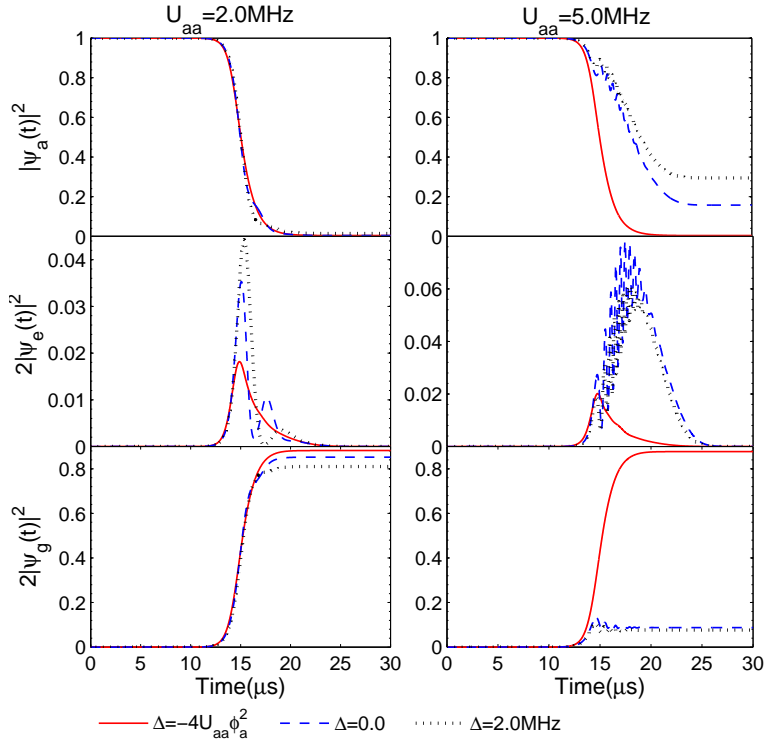}%
\caption{Time evolution of the populations for ground-state atoms $|\psi_{a}(t)|^{2}$, excited-state molecule 2$|\psi_{e}(t)|^{2}$ and ground-state molecules 2$|\psi_{g}(t)|^{2}$. Other parameters $\Omega_{p,s}^{0}$, $t_{p,s}$ and $\tau$ are the same as used in figure \ref{stability_dia} }.%
\label{population_opt}%
\end{figure}

Now we proceed to study how the atom-atom collision and the effective one-photon resonance could affect the atom-molecule transfer processing. Figure \ref{population_opt} shows the time evolution of ground-state atom population(upper), excited-state molecule population(middle) and ground-state molecule population(bottom) for one case with an effective one-photon resonance ($\Delta_{u}=0$, solid), and for other two cases with $\Delta=0$(dashed) and 2.0MHz(dotted). A state very close to the CPT state ($|\psi_{e}|^{2}\approx0$) is almost produced for different detuning schemes. But, since a good adiabatic condition of our CPT state could be achieved with effective one-photon detuning even in dynamical unstable regions, figure \ref{population_opt} shows that the excited molecule population for the case with effective one-photon is much lower than those without this condition. 
A good adiabaticity could guarantee a high atom-molecule transfer efficiency when the atom-atom collision presents. The left and right panels of figure \ref{population_opt} plot a comparison of the atom-molecule transfer processing for two quite different atom-atom collisional strengths $U_{aa}$=2.0MHz and 5.0MHz. When $U_{aa}$ is relatively low (2.0MHz), the final atom-molecule transfer efficiency is not sensitive to single photon detuning. In contrast, when $U_{aa}$ increases to 5.0MHz,  the final transfer efficiency becomes very sensitive to it: without effective single-photon resonance, the final efficiency is just around 10$\%$, and a robust atom-molecule transfer (efficiency$>80$$\%$) can be achieved when the effective single photon detuning is employed. In fact, the CPT state in our scheme already includes the collisional effect. When the collision strength $U_{aa}$ is relatively weak, the CPT state with collisions is almost close to those without collisions. When the collisional strength is strong enough, collision effect has strong influences on the stability of the CPT state and thus the final atom-molecule transfer efficiency becomes sensitive to the detuning. Fortunately, the collision nonlinearity could be compensated by using the effective single photon resonance and thus we can still achieve a much higher atom-molecule transfer efficiency than those not satisfying this condition.    

\subsection{Experimental Feasibility}

Before ending, it is necessary for us to verify the experimental feasibility of
our scheme, especially the parameters we have used in the calculations. Table \ref{table1} organized from \cite{Julienne06} enables to show the background $s$-wave
scattering lengths $a_{bg}$ and collisional strengths $U_{aa}$ of some usual
alkali-metal atoms.
\begin{table}[hm]
\caption{Parameters characterizing the background scattering length $a_{bg}$
associated with the $s$-wave collisional potentials $U_{aa}$. $n_{0}$ denotes
the initial condensate density, being $n_{0}=10^{21}$m$^{-3}$, $m$ is the
atomic mass.}%
\label{table1}
\begin{indented}
\item[]\begin{tabular}{@{}l*{15}{c}}
\br
Species & $B_{0}($G$)$ & $a_{bg}($nm$)$ & $U_{aa}=\frac{4\pi\hbar
a_{bg}n_{0}}{m}$(MHz)\\
\mr
$^{6}$ Li & 543.25(5) & 3.122 & 0.415\\
& 834.149 & -74.348 & -9.880\\
$^{23}$ Na & 853 & 3.381 & 0.117\\
& 907 & 3.323 & 0.115\\
$^{40}$ K & 202.10(7) & 9.208 & 0.184\\
& 224.21(5) & 9.208 & 0.184\\
$^{85}$ Rb & 155.0 & -23.442 & -0.220\\
$^{87}$ Rb & $ {1007.40(4)}$ & 5.318 & 0.049\\
$^{133}$ Cs & 19.90(3) & 8.625 & 0.052\\
& 47.97(3) & 47.890 & 0.287\\\hline\hline
\end{tabular}
\end{indented}
\end{table}

Table \ref{table1} provides the parameters $a_{bg}$ and $U_{aa}$
characterizing the background scattering potential for experimentally relevant
zero-energy resonances. Explicitly, almost all the collisional strengths
$\left\vert U_{aa}\right\vert $ values are much smaller than 10.0MHz(except
$^{6}$Li at $B_{0}=834.149$G). Thereby, results discovered from figures \ref{population_dy} and \ref{population_opt} are universal and can be available with any atomic species.
In our other numerical simulations (not shown), we use laser pulses with different intensities and widths, and even take other collisions into account, including collisions between atom-molecule and molecule-molecule.
Pleasantly, we find that the results are still
perfectly consistent with figure \ref{population_dy} and other texts.

Finally, we note that the effective one- and two-photon resonances, which are the keys to obatin a high atom-molecule transfer efficiency, can be realized with current technique of frequency chirped pump and control laser pulses \cite{Cheng06,Vitanov11}.  In experiments, it's hard to exactly maintain the effective single photon resonance. Actually, Figure \ref{adiabaticity_pic} also shows that when $|\Delta_{u}|\leq2.0$MHz, $r$ is much smaller than 0.1, very good adiabaticity and a high atom-molecule transfer efficiency could be obtained.

\section{Summary}

In summary, we have investigated the atom-molecule transfer within a $\Lambda$-type configuration including atom-atom collision. We first obtain a CPT state including the atom-atom collisional effect under the effective two-photon resonance condition and find the collision induced frequency shift between two quantum states. Further, we study the stability and adiabaticity of the CPT state. We find that instability could be essentially induced when the pump and control lasers overlap and thus when the overlapping time of the pump and control lasers is short so that the unstable growth of the deviation from the CPT state could be neglected,  good adiabaticity of the CPT state can be obtained in some unstable regions, not just in stable ones. Employing the effective single photon resonance, our numerical simulations show that a rather high atom-molecule transfer efficiency ($>$80$\%$) could be achieved. Finally, the feasibility of our scheme for future experiments is discussed. 
Our findings may provide a new way to efficiently photoassociate ground state molecules from a cold atomic gas with strong atom-atom collisional interaction.

%

\ack
This work was supported by the National Basic Research Program of China (973 Program) under Grants No. 2011CB921602 and No. 2011CB921604, and the National Natural Science Foundation of China under Grant No. 11104076, No. 11034002 and No. 10874045 and the Specialized Research Fund for the Doctoral Program of Higher Education no 20110076120004.

\end{document}